\documentclass[prb,10pt]{revtex4-1}

\usepackage{amsmath}
\usepackage{amsfonts}
\usepackage{amssymb}
\usepackage{graphicx}
\usepackage{mathrsfs}
\usepackage{times}
\usepackage{color}
\usepackage{psfrag}
\usepackage{natbib}
\definecolor{violet}{rgb}{0.5,0,0.5}
\definecolor{blue2}{rgb}{0.5,0,1}
\usepackage{setspace}
\usepackage{graphicx,subfig}
\usepackage{array,calc}
\usepackage{braket}
\usepackage{bm}
\usepackage[justification=raggedright]{caption}

\begin{document}


\title{Space-time wave localization in electromechanical metamaterial beams with programmable defects}

\author{Renan Lima Thomes$^1$} 
\author{Danilo Beli$^1$} 
\author{Carlos De Marqui Jr.$^1$} 

\affiliation{$^1$S\~{a}o Carlos School of Engineering, University of S\~{a}o Paulo, Brazil}




\date{\today}

\begin{abstract}
A flexible wave localization is investigated using a spatial-temporal modulation of point defects along the periodic array of electromechanical local resonators of a piezoelectric bimorph beam. By changing the electrical resonance of subsequent unit cells through a smooth and synchronized strategy, a point defect is moved along the metastructure, which properly transfers the vibration energy between the unit cells as well as perfectly induces wave localization according to the final defect position. This wave manipulation by programmable defects is also applied to a two-dimensional triangular lattice composed by piezoelectric metamaterial beams, which illustrates the potential and versatility of the proposed strategy in more complex systems. Therefore, this work opens new possibilities for programmable wave localization and manipulation using smart electromechanical metamaterials with applications that may involve guiding, sensing, monitoring and harvesting.
\end{abstract}





\maketitle

\section{Introduction}

Phononic crystals \cite{Sigalas1992a} and dynamic metamaterials (e.g., elastic and acoustic) \cite{Liu2000} are engineered structures with the ability to modify the wave propagation through band gaps, i.e. frequency bands where waves cannot propagate. The band gaps of phononic crystals open due to periodic mismatch of properties (e.g., geometry and material) at wavelengths comparable to the spatial periodicity, in which incident and reflect waves cancel each other leading to the Bragg scattering \cite{Sigalas1992a}. Therefore, low-frequency Bragg band gaps are possible only in large structures \cite{Martinez-Sala1995,Sugino2017a}. On the other hand, metamaterials open band gaps due to arrays of local resonating attachments which provide interesting and unusual dynamic properties at low frequencies \cite{Liu2000,Hussein2014}. Due to their unique wave manipulation capabilities, phononic crystals and metamaterials have been used for a wide variety of applications over the past two decades, such as acoustic and vibration isolation \cite{Sanchez-Perez1998,Ho2003a,Bilal2013}, cloaking \cite{Cummer2007,Norris2011}, focusing \cite{Guenneau2007}, sound collimation \cite{Chen2004,Christensen2007} and refraction of sound waves \cite{Cervera2001,Zhang2004,Hussein2014}.

An important feature for various applications of engineered structures is the wave localization and guidance using defects \cite{Sigalas1997,Sigalas1998,Torres1999,Kafesaki2001,Khelif2003}. Point defects can be introduced in phononic crystals by replacing a unit cell by another one with different geometric or material properties \cite{Sigalas1997}. In metamaterials, a point defect is created by replacing a local resonator with another one that has significantly different dynamic properties \cite{Kaina2017}. By introducing a point defect, the local periodicity is broken and a flat pass band (or defect band) emerges within the band gap \cite{Sigalas1997,Shakeri2019}. Therefore, energy can be confined inside and at the vicinity of the defect, resulting in wave localization. In addition, line defects are created when a chain of unit cells is replaced by defects, resulting in a waveguide where elastic wave propagation is confined by the band gaps of adjacent unit cells \cite{Sigalas1998}. Once the topology or material properties of phononic crystals are hardly reconfigurable \cite{Wang2020}, or if one cannot control the dynamic properties of resonant attachments of metamaterials, the elastic wave manipulation through defects in periodic systems would be limited to fixed configurations \cite{Liu2000,Martinez-Sala1995,Beli2019, Wang2020}. Therefore, efforts towards tunable architected materials have been reported in the literature \cite{Wang2020}, paving the way for remarkable applications such as tunable wave localization \cite{Hu2014,Shakeri2019}; reconfigurable wave guiding \cite{Wang2017, Jin2016, Casadei2012,HwanOh2011,Li2019} with filtering \cite{Khelif2003b} and (de)multiplexing properties \cite{Pennec2004,Mohammadi2011,Rostami-Dogolsara2016,Vasseur2011,Wang2017}; as well as new concepts on vibration-based energy harvesting \cite{Carrara2013,Oudich2017,Jo2020,Jo2021}.

Tunable or programmable phononic crystals and metamaterials are usually obtained by mechanical or multifield coupling-based reconfigurability \cite{Wang2020}. For instance, reconfigurable defects, and consequently tunable wave localization, have been created in a square-lattice phononic crystal composed of periodic circular holes embedded in a metallic matrix in which some selected holes are filled with liquid \cite{Wang2017}. Furthermore, the transmission of the confined waves through reconfigurable circuits (waveguides) along a metallic matrix can be performed due to the tunable fluid filled defects \cite{Jin2016}. Among different possibilities for multifield coupling, piezoelectric materials are a relatively popular choice \cite{Thorp2001,Wu2009,Spadoni2009,Airoldi2011,Casadei2012,Casadei2010,Wang2014,HwanOh2011,Shakeri2019,Li2019,Yi2019,Sugino2020a,Sugino2020,Silva2020}. In such case, the wave behavior of electromechanically coupled metamaterials and phononic crystals can be controlled by updating external passive shunt circuits or active circuits \cite{Wu2009,Wang2014,Shakeri2019,Thorp2001,Airoldi2011,Spadoni2009,Casadei2010,Yi2019, Sugino2020a}. Electromechanically coupled phononic crystals can realize arbitrarily (reconfigurable) shaped waveguides by short-circuiting selected paths of piezoelectric inclusions while leaving the remaining units in open-circuit condition \cite{HwanOh2011}. Bragg scattering band gap of a plate with periodic array of cylindrical stubs combined with piezoelectric resonators allows the control of elastic waves propagation through pre-configured waveguides of the phononic crystal \cite{Casadei2012}. Piezoelectric material connected to negative capacitance circuits can also tune the propagation direction in relatively complex shaped waveguides of elastic metamaterial plates \cite{Li2019}, realizing switchable waveguides. Recently, the use of digitally controlled analog circuits \cite{Silva2020} and digital shunt circuits \cite{Sugino2020} replacing analog shunts has facilitated experimental implementations of reconfigurable matamaterials. In particular, practical implementation of programmable metamaterials using digital circuits has been recently presented in literature \cite{Sugino2020}.

In this work, we present a novel programmable piezoelectric metamaterial that can confine elastic energy and perform space-time wave localization by programming defects in a periodic array of electromechanical resonators. The manipulation of localized waves in space and time is induced by strategically controlling the inductance of shunt circuits (i.e., local resonators) connected to the electromechanical unit cells. In Section 2, this concept is presented using a piezoelectric metamaterial beam, a reconfigurable system that allows universal wave localization through space and time. The numerical model of the system is also presented in this section. The behavior of this metamaterial with a fixed defect is analyzed in Section 3 through its band structure, harmonic response and time response. Later, in the same section, the tunable strategy that modulates the defects is presented and the space-time wave localization by programmable defects is discussed. A more complex example is shown in Section 4, where the space-time wave localization is realized in a two-dimensional triangular lattice composed by the combination of piezoelectric metamaterial beams. Finally, the concluding remarks are discussed in Section 5.

\section{Electromechanical metamaterial model}\label{s:model}
\begin{figure}[hbt]
	\centering
	\includegraphics[scale = 0.85]{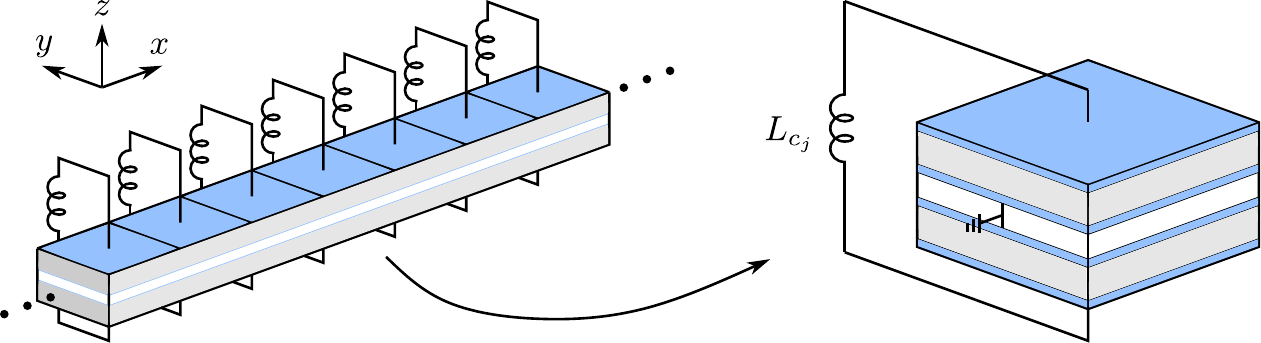}
	\caption{Electromechanical metamaterial beam in a bimorph configuration  (left) and its unit cell in detail (right). The colored regions correspond to the substructure (white), the piezoelectric material (gray), the electrodes (blue) and the shunt circuits composed by inductors $L_c$ (black).}
	\label{f:beam}
\end{figure}

Consider an elastic beam composed of a conductive substructure bracketed by two piezoelectric layers oppositely poled in the $z$ direction (i.e., a bimorph configuration) as in Fig. \ref{f:beam}. The piezoelectric layers are covered on top and bottom surfaces by thin layers of conductive electrodes that are periodically segmented and connected in series to inductive shunt circuits, which are the local resonators. The boundaries of each segmented electrode region delimit the unit cell of the periodic electromechanical metamaterial depicted in Fig. \ref{f:beam}. While the unit cells are electrically independent from each other, the right boundary of the $j$-th unit cell and the left boundary of the unit cell $j+1$ share the same mechanical degrees of freedom (DOFs).

 Since the analyses in this work are limited to thin beams, the Euler-Bernoulli assumptions are employed in the formulation of an electromechanically coupled finite element (FE) model of the metamaterial beam. The continuum is discretized into one-dimensional beam elements, each one containing two nodes at its boundaries. The transverse displacements within each element are assumed to be a cubic interpolation of its nodal transverse displacements ($w$) and rotations about the $y$ direction ($\theta$) \cite{Petyt1990}. In the model derivation, the non-zero electric field component ($z$ direction) is assumed uniform in the thickness direction of the piezoelectric element and the effect of structural damping is not taken into account. Therefore, the dynamic behavior of the electromechanical metamaterial beam can be described by the following finite element equations \cite{Carpenter1997,DeMarquiJunior2009}:
\begin{subequations}\label{e:before_final}
	\begin{align}
	& \bf{M}\ \ddot{\bf{u}} + \bf{K}\ \bf{u} - \bf{\Theta}\ \bf{v} = \bf{F} \\ 
	& \bf{C_p}\ \bf{v} +\bf{q} + \bf{\Theta}^t\ \bf{u} = \bf{0} \, ,
	\end{align}
\end{subequations}
where $\bf{M}$ is the global mass matrix $(m\times m)$, $\bf{K}$ is the global stiffness matrix $(m\times m)$,  $\bf{\Theta}$ is the electromechanical coupling matrix $(m\times e)$, $\bf{C_p}$ is the diagonal capacitance matrix $(e\times e)$, $\bf{F}$ is the vector of external mechanical forces $(m\times 1)$, $\bf{v}$ is the vector of voltage output from each unit cell $(e\times 1)$, $\bf{q}$ is the vector of electrical charges ($e\times 1$), and $\bf{u}$ is the vector of mechanical DOFs $(m\times 1)$, which contains the transverse displacements and the rotations (about the $y$ direction) associated to each node. Here, $m$ and $e$ refer to the number of mechanical and electrical DOFs, respectively. The over-dots denote time differentiation and $t$ denotes the matrix transpose when used as a superscript, otherwise, it stands for time. The inductive shunt circuits with time-varying inductances connected to the unit cells are modeled as
\begin{equation}\label{key}
\bf{v} = \frac{\mathrm{d}}{\mathrm{d}t} \left( \bf{L_c} \frac{\mathrm{d}\bf{q}}{\mathrm{d}t} \right)\, ,
\end{equation}
where $\bf{L_c}$ is a $(e\times e)$ diagonal matrix containing the inductances of each shunt circuit. For systems with time-varying inductances considered in Sections \ref{ss:time}-\ref{s:lattice}, the matrix representation of Eq. \ref{e:before_final} is
\begin{equation}\label{e:gov-q}
\begin{bmatrix}
\bf{M} & -\bf{\Theta}\, \bf{L_c}\\
\bf{0} & \bf{C_p}\, \bf{L_c}
\end{bmatrix}
\begin{bmatrix}
\ddot{\bf{u}}\\
\ddot{\bf{q}}
\end{bmatrix} +
\begin{bmatrix}
\bf{0} & -\bf{\Theta}\, \dot{\bf{L_c}}\\
\bf{0} & \bf{C_p}\, \dot{\bf{L_c}}
\end{bmatrix}
\begin{bmatrix}
\dot{\bf{u}}\\
\dot{\bf{q}}
\end{bmatrix} +
\begin{bmatrix}
\bf{K} & \bf{0}\\
\bf{\Theta}^t & \bf{I}
\end{bmatrix}
\begin{bmatrix}
\bf{u}\\
\bf{q}
\end{bmatrix} =
\begin{bmatrix}
\bf{F}\\
\bf{0}
\end{bmatrix},
\end{equation}

where $\bf{I}$ is the identity matrix. When the time variation of $\bf{L_c}$ is negligible, Eq. \ref{e:before_final} reduces to:
\begin{equation}\label{e:gov-ns}
\begin{bmatrix}
\bf{M} & \bf{0}\\
\bf{\Theta}^t & \bf{C_p}
\end{bmatrix}
\begin{bmatrix}
\ddot{\bf{u}}\\
\ddot{\bf{v}}
\end{bmatrix} +
\begin{bmatrix}
\bf{K} & -\bf{\Theta}\\
\bf{0} & \bf{L_c}^{-1}
\end{bmatrix}
\begin{bmatrix}
\bf{u}\\
\bf{v}
\end{bmatrix} =
\begin{bmatrix}
\bf{F}\\
\bf{0}
\end{bmatrix}.
\end{equation}

The system of equations in Eqs. \ref{e:gov-q}-\ref{e:gov-ns} characterizes a multiphysics problem in which the mechanical and electrical domains are coupled by the off-diagonal terms of the matrices, i.e., the electromechanical coupling matrix $\bf{\Theta}$. For the $j$-th unit cell, the inductive shunt combined with the effective capacitance of the bimorph results in a capacitive-inductive circuit whose resonant frequency is \cite{Hagood1991}
\begin{equation}\label{key}
\omega_{n_j} = 1/\sqrt{C_{p_j} L_{c_j}}.
\end{equation}
where $C_{p_j}$ and $L_{c_j}$ are the capacitance of the $j$-th unit cell and the inductance of the $j$-th shunt circuit, respectively. Conversely, an electrical resonance can be tuned to a selected frequency by setting its inductance to $L_{c_j} = 1/(C_{p_j}\, \omega_{n_j}^2)$. 

\subsection{Dispersion computation}
The spectral properties of a periodic structure can be obtained by applying Bloch-Floquet conditions at the unit cell boundaries and solving the resulting eigenvalue problem. The unit cell, or a set of unit cells (that results in a supercell, as will be later discussed in Section \ref{ss:spectral}) can be modeled using a FE model \cite{Orris1974}, such as the one briefly discussed in Section \ref{s:model} for electromechanically coupled media (Eq. \ref{f:beam}) that results in Eq. \ref{e:gov-ns}. However, matrix symmetry is highly recommended for solving such eigenvalue problems \cite{Everstine1981} and, therefore, Eq. \ref{e:gov-ns} is modified by time integrating its second line as well as multiplying the first line by -1, resulting in the symmetric matrix form,

\begin{equation}\label{e:gov-sym}
    \tilde{\bf{M}}\, \ddot{\bf{p}} + \tilde{\bf{C}}\, \dot{\bf{p}} + \tilde{\bf{K}}\, \bf{p} = \tilde{\bf{F}}
\end{equation}
with
\begin{equation}\label{e:tildes}
\tilde{\bf{M}} =  
\begin{bmatrix}
-\bf{M} & \bf{0}\\
\bf{0} & \bf{C_p}
\end{bmatrix}, \ 
\tilde{\bf{C}} = 
\begin{bmatrix}
\bf{0} & \bf{\Theta}\\
\bf{\Theta}^t & \bf{0}
\end{bmatrix}, \
\tilde{\bf{K}} = 
\begin{bmatrix}
\bf{-K} & \bf{0}\\
\bf{0} & \bf{L_c}^{-1}
\end{bmatrix}, \
\tilde{\bf{F}} = 
\begin{bmatrix}
\bf{-F}\\
\bf{0}
\end{bmatrix} \text{ and }
\bf{p} = 
\begin{bmatrix}
\bf{u}\\
\bf{\xi}
\end{bmatrix}.
\end{equation}
where $\bf{\xi}$ is an auxiliary variable with $\dot{\bf{\xi}}=\bf{v}$. By assuming harmonic motion, spectral solutions can be imposed: $\bf{p}(t) = \bar{\bf{p}}(\omega)\mathrm{e}^{\mathrm{i}\omega t}$ and $\tilde{\bf{F}}(t) = \bar{\bf{F}}(\omega)\mathrm{e}^{\mathrm{i}\omega t}$, where $\omega$ is the angular frequency and  $\mathrm{i}$ is the unit imaginary number. Accordingly, Eq. \ref{e:gov-sym} is transformed from time to frequency 
\begin{equation}\label{e:orig}
(-\omega^2\tilde{\bf{M}} + \mathrm{i}\omega\tilde{\bf{C}}  + \tilde{\bf{K}} )\, \bar{\bf{p}} = \bar{\bf{F}}.
\end{equation}
Since the structure is periodic, the degrees of freedom from the left and right unit cell boundaries can be related by the Bloch-Foquet periodic condition (i.e., $\bar{\bf{p}}_{R} = \bf{I}e^\mu \bar{\bf{p}}_{L}$), which transforms the physical space in the reciprocal space (i.e., wavenumber domain). Therefore, the complete set of DOFs ${\bar{\bf{p}}}$ are related to the reduced set of DOFs $\hat{\bf{p}}$ \cite{Hussein2014} by 
\begin{equation}\label{e:reduc}
\bar{\bf{p}} = \bf{T}\, \hat{\bf{p}}, \quad \text{with} \quad
\bar{\bf{p}} = 
\begin{bmatrix}
\bar{\bf{p}_I}\\
\bar{\bf{p}_L}\\
\bar{\bf{p}_R}
\end{bmatrix}, \quad \bf{T} =
\begin{bmatrix}
\bf{I} & \bf{0}\\
\bf{0} & \bf{I}\\
\bf{0} & \bf{I}e^\mu
\end{bmatrix}, \quad \hat{\bf{p}} =
\begin{bmatrix}
\bar{\bf{p}_I}\\
\bar{\bf{p}_L}\\
\end{bmatrix},
\end{equation}
where $\bf{T}$ denotes a linear transformation and the subscripts $I$, $L$ and $R$ stand for the interior, left and right DOFs, respectively. The voltage output of each unit cell is settled as interior DOFs. Additionally, $\mu=-\mathrm{i}\, \kappa\, \Delta$ is the propagation constant, where $\kappa$ is the wavenumber and $\Delta$ is the length of the considered periodicity (i.e., for a unit cell $\Delta = a$ and for a supercell $\Delta = na$, where $a$ is the unit cell length and $n$ is the number of unit cells). In the absence of external forces on the interior nodes, the equilibrium at the left DOFs implies that the sum of nodal forces of the elements connected to the left DOFs is zero \cite{Mace2008}, $\bf{\bar{T}}^t\, \bf{f} = \bf{0}$, where
\begin{equation}\label{key}
\bf{\bar{T}}^t = 
\begin{bmatrix}
\bf{I} & \bf{0} & \bf{0}\\
\bf{0} & \bf{I} & \bf{I}e^{-\mu}
\end{bmatrix}.
\end{equation}
Substituting Eq. \ref{e:reduc} into Eq. \ref{e:orig} and premultiplying it by $\bf{\bar{T}}^t$ yields the eigenvalue problem $\omega(\kappa)$
\begin{equation}\label{e:key}
(-\omega^2 \hat{\bf{M}} + \mathrm{i}\omega \hat{\bf{C}} + \hat{\bf{K}})\hat{\bf{p}}=\bf{0},
\end{equation}
with the reduced matrices $\hat{\bf{M}}(\kappa)=\bf{\bar{T}}^t\tilde{\bf{M}}\bf{T}$, $\hat{\bf{C}}(\kappa)=\bf{\bar{T}}^t\tilde{\bf{C}}\bf{T}$ and $\hat{\bf{K}}(\kappa)=\bf{\bar{T}}^t\tilde{\bf{K}}\bf{T}$. Finally, the eigenvalue problem is solved by sweeping through all values of $\kappa$ along the first irreducible Brillouin zone \cite{Hussein2014}, i.e., $\kappa \in [0\ \pi/\Delta] $, thus finding the eigenvalues $\omega_j$ and eigenvectors (or wave mode shapes) $\hat{\bf{p}}_j$ as a function of $\kappa_j$. $\hat{\bf{p}}_j$ depicts the spatial distribution of the DOFs, whereas $\kappa_j$ is the phase change per unit length \cite{Mace2005,Beli2019}. 
\subsection{Material and geometric properties of the unit cell}

Each unit cell is composed of an aluminum substructure bracketed by a pair of PMN-PT piezoelectric elements poled oppositely in the $z$ direction and connected in series to a resonant shunt circuit. The PMN-PT single crystal was chosen due to its high electromechanical coupling, which leads to wider band gaps \cite{Sugino2017} and renders the studied phenomena more evident. The material properties are presented in Table \ref{t:params}. The length and width of each unit cell are 12.7 mm, the thickness of each piezoelectric layer is 0.25 mm while the substructure is 0.167-millimeter thick.

\begin{table}[hbt]
	\begin{center}
		\caption{Material properties for the aluminum substructure and the PMN-PT.}
		\label{t:params}
\begin{tabular}{l l}
\hline
\textbf{Property}  & \textbf{Value} \\ 
\hline
Young's modulus of the substructure, $Y_s$ & 69 GPa  \\
				Density of the substructure, $\rho_s$                                         & 2700 $\mathrm{kg/m}^3$                    \\
				Density of the piezoelectric material, $\rho_p$                                & 8120 $\mathrm{kg/m}^3$                    \\
				Elastic compliance at constant electric field, $s_{11}^E$ 					   & $45.9 \times 10^{-12}\ \mathrm{Pa}^{-1}$ \\
				Piezoelectric strain coefficient, $d_{31}$               					   & $-646\times 10^{-12}$ V/m       \\
				Piezoelectric permittivity constant at constant stress, $\varepsilon_{33}^T$ \quad  \quad \quad  & $4.208\times 10^{-8}$ F/m     
\end{tabular}
{\small \item The subscript numbers refer to the Voigt notation of the tensor, where 1, 2 and 3 corresponds to the $x$, $y$ and $z$ directions, respectively.}
	\end{center}
\end{table}

Throughout this text, the frequencies are normalized by the velocity of the longitudinal wave considering the average properties of the composite structure. Therefore, the normalized frequency is $\Omega = \omega\, \Delta/\left( 2\pi\sqrt{Y_{avg}/\rho_{avg}}\right) $, where $Y_{avg}$ is the average Young's modulus and $\rho_{avg}$ is the average density given by
\begin{equation}\label{key}
1/Y_{avg} = v_f/Y_{s} + (1-v_f)/Y_{p}\, \text{, and}\
\end{equation}
\begin{equation}\label{key}
\rho_{avg} = v_f\, \rho_s + (1-v_f)\, \rho_p,
\end{equation}
where $v_f$ is the volume fraction of the substructure with respect to the total volume of the system ($v_f = 0.2504$), and $Y_p=1/s_{11}^E$. Each unit cell is discretized into eight finite elements, such that the length of each element corresponds to 4.6\% of the smallest wavelength analyzed, which is sufficient to properly capture the wave propagation phenomenon.


\section{Piezoelectric metamaterial beam with defects}\label{s:beam}
In this section, the dynamic behavior of the metamaterial beam with a defect (fixed in space and time) is first discussed considering spectral analysis and  harmonic response. Later, the time domain response is analyzed and the space-time wave localization by programmable defects is discussed in detail.


\subsection{Spectral and harmonic analyses of a fixed defect}\label{ss:spectral}

\begin{figure}[hbt]
	\centering
	\includegraphics[]{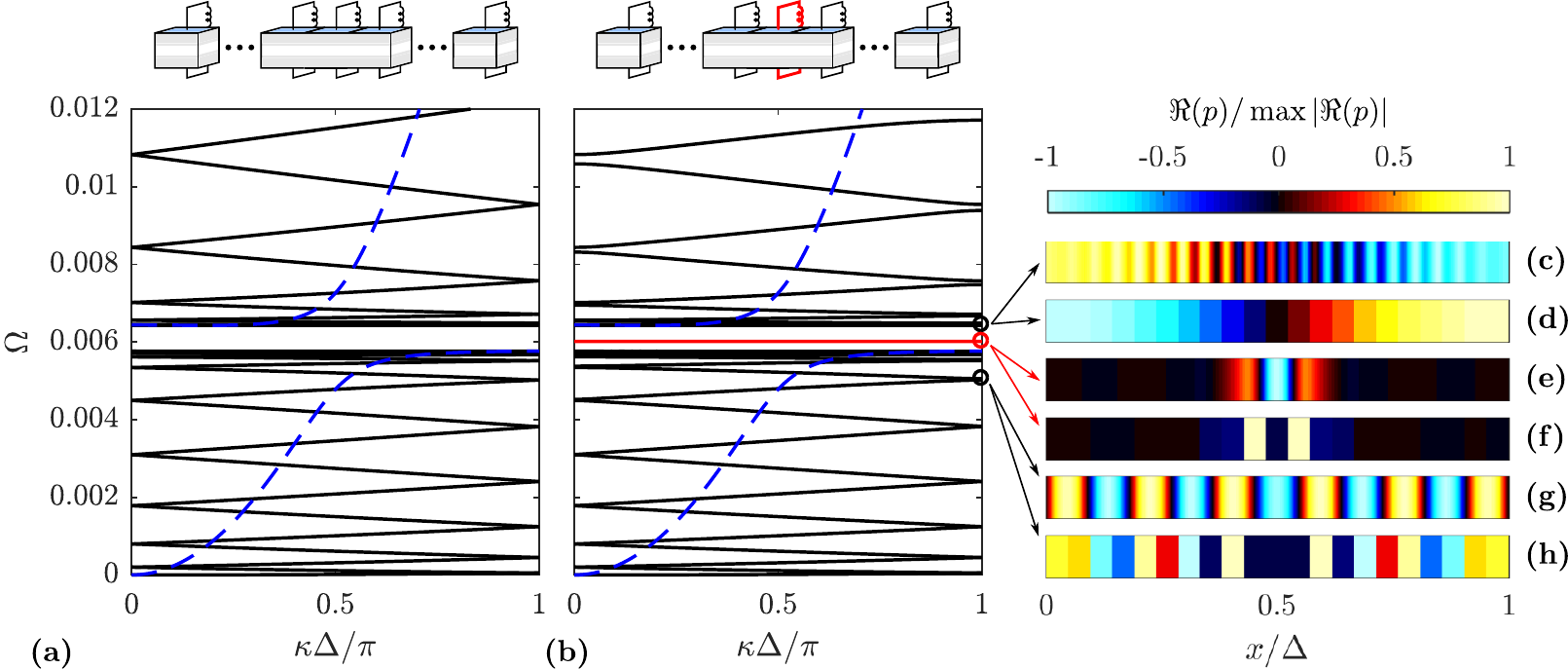}
	\caption{Band structure of the periodic metamaterial (a) and metamaterial with defect (b): single unit cell $\Delta = a$ (blue dashed lines) and supercell with $\Delta = na$ (black and red lines). The displacement field (c, e, g) and the correspondent voltage field (d, f, g) for the wave mode shapes around the band gap (c-d and g-h) and at the defect band (e-f). The defect wave mode is depicted in red at the band structure (b).}
	\label{f:bs}
\end{figure}

First, the band structure of the piezoelectric metamaterial beam without and with a point defect is computed considering a supercell composed of 21 unit cells (i.e., $n = 21$). This procedure, where the band structure is calculated using several unit cells, is known as supercell technique \cite{Sigalas1997}, in which the same equation for one unit cell, Eq. \ref{e:key}, can be used for the dispersion computation of the supercell by changing only the correspondent finite element matrices and the periodicity length $\Delta$. Fig. \ref{f:bs}(a) displays the band structure of the periodic metamaterial supercell (without defect) with the resonant frequency of each local resonator tuned to $\Omega_t = 6.44\times10^{-3}$ ($\Omega_{n_j} = \Omega_t\ \forall\ j$), which is referenced throughout this work as the tuning frequency. The resulting locally resonant band gap has a lower limit at $\Omega_l= 5.76\times10^{-3}$ and an upper limit at $\Omega_u = 6.44\times10^{-3}$. Moreover, the band structure for a single unit cell is superimposed in the figure as a reference (blue dashed line), both computations show the same frequency zones for the bulk bands and the band gap. Figure \ref{f:bs}(b) presents the band structure for the piezoelectric metamaterial beam with a defect introduced at its center (i.e., 11\textsuperscript{th} unit cell). This defect is created by changing the inductance of the 11\textsuperscript{th} electromechanical local resonator in order to tune its frequency to $2\Omega_t$ (i.e., $\Omega_{n_{11}} = 2\Omega_t$).  Therefore, the resonant shunt of the defected unit cell works like an open circuit compared to the behavior of the other unit cells at frequencies around $\Omega_t$. As a consequence, a defect band (the red line in Fig. \ref{f:bs}(b)) is created within the band gap at $\Omega =6.01\times 10^{-3}$, which is the defect-mode frequency. Since the defect band is flat (i.e., nearly zero group velocity), the energy can be localized and confined inside and in the vicinity of this defected unit cell at the defect frequency. Please, check Appendix \ref{ap:detuning} for a more detailed discussion regarding the effects of different electromechanical resonaces on the defect properties. 

Figure \ref{f:bs}(c-h) displays the wave modes related to transverse displacements (Fig. \ref{f:bs}(c, e, g)) and voltage outputs (Fig. \ref{f:bs}(d, f, h)) calculated at $\kappa\Delta/\pi = 1$ for the metamaterial with a point defect. At the band gap limits (e.g., Fig. \ref{f:bs}(c, d)) and at the propagating bands (e.g., Fig. \ref{f:bs}(g, h)), the vibration energy is evenly distributed along the supercell. On the other hand, for the wave modes at the defect-band frequency (Fig. \ref{f:bs}(e, f)), the vibration energy is localized inside and in the vicinity of the defect, confirming a localized defect mode. Moreover, the wave modes are anti-symmetric in Fig. \ref{f:bs}(c, d), whereas they are symmetric in Fig. \ref{f:bs}(e-h). Although the displacement field in Fig. \ref{f:bs}(g) has a periodic pattern that repeats approximately every $x/\Delta=2/11$, the correspondent voltage field in Fig. \ref{f:bs}(h) does not exhibit the same behavior since the electrodes are segmented every $x/\Delta=1/21$ (which is not a factor of 2/11). Moreover, the FE furnish a displacement distribution along the beam, while a single voltage output is obtained for each unit cell.

\begin{figure}[h!]
	\centering
	\includegraphics[]{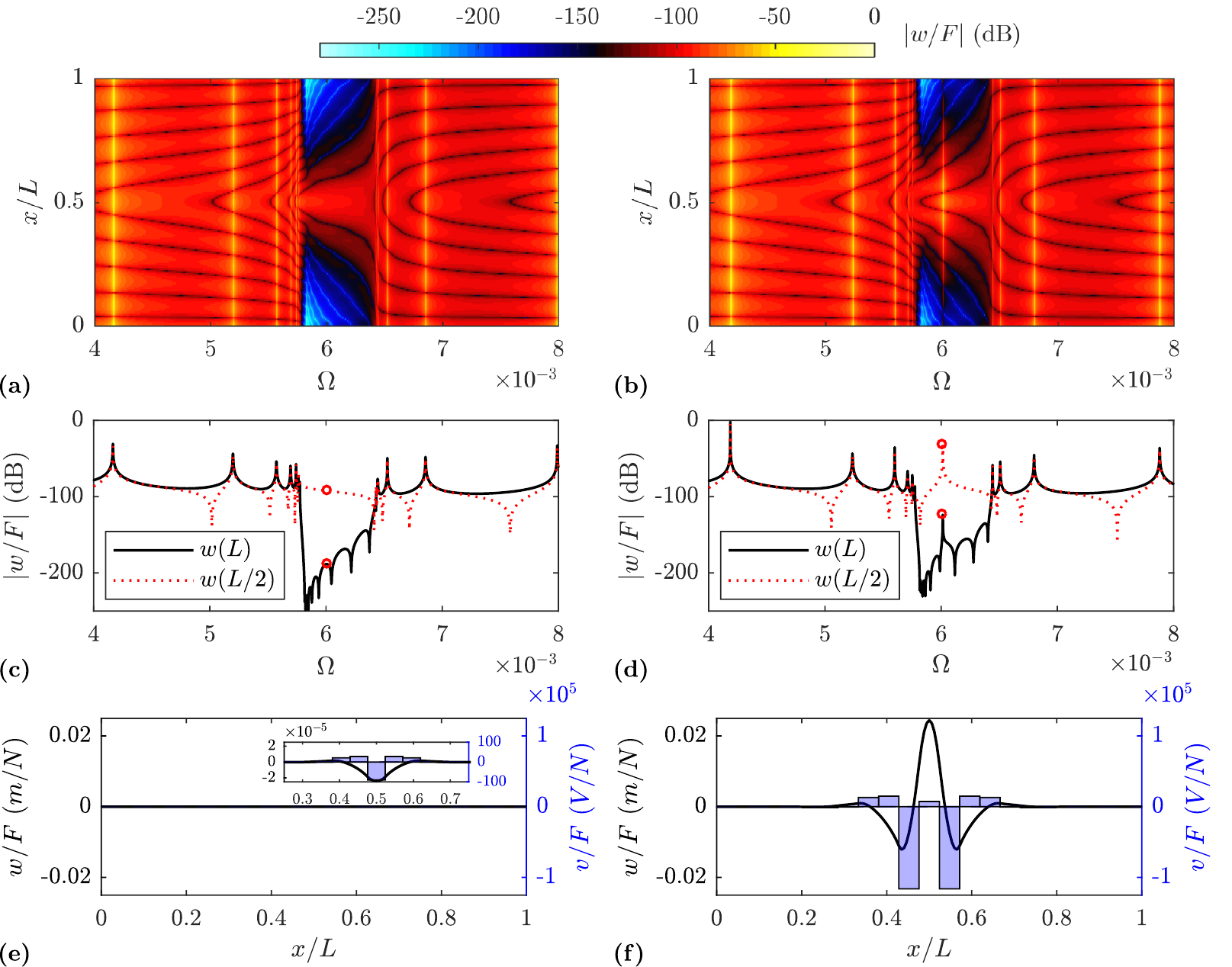}
	\caption{Maps of FRFs as a function of the space (a, b). FRF at the middle and at the end of the metastructure (c, d). Displacement and voltage fields ($w(x)$ and $v(x)$) at the defect-mode frequency indicated by the circles in literals c and d (e, f). Results for the periodic metastructure (a, c, e) and for the metastructure with defect (b, d, f). Moreover, $L$ represents the total metastructure length.}
	\label{f:harmonic}
\end{figure}

The harmonic responses of the previous piezoelectric metamaterial beams composed of 21 unit cells are also investigated. For this purpose, an harmonic punctual excitation $F(x_{11},\Omega)$ is applied at the middle of the 11\textsuperscript{th} unit cell and the correspondent transverse displacements $w(x,\Omega)$ are computed along the beam using Eq. \ref{e:gov-ns}. \ref{f:harmonic}(a) displays the frequency response function (FRF), $|w(x,\Omega)/F(x_{11},\Omega)|$, along the periodic metamaterial beam with the shunt circuits tuned to $\Omega_t = 6.44\times10^{-3}$. A vibration attenuation zone is observed at the frequency range of the locally resonant band gap shown in Fig. \ref{f:bs}(a). When a defect is introduced at the 11\textsuperscript{th} local resonator such that $\Omega_{n_{11}} = 2\Omega_t$, a mechanical resonance, and hence, a vibration energy localization emerges within the band gap at $\Omega=6.01\times 10^{-3}$ (Fig. \ref{f:harmonic}(b)), which corresponds to the dynamic behavior of the defect-band observed in Fig. \ref{f:bs}(b). Figure \ref{f:harmonic}(c, d) displays the FRFs evaluated at the middle and at the end of the metamaterial beams. The defected metastructure (Fig. \ref{f:harmonic}(d)) exhibits a vibration mode with high amplitude inside the vibration attenuation zone when compared to correspondent metastructure without defect (Fig. \ref{f:harmonic}(c)). 

The displacement and voltage fields at the defect-band frequency are presented in Fig. \ref{f:harmonic}(e, f). The vibration amplitudes of the periodic configuration (Fig. \ref{f:harmonic}(e)) are negligible compared to the amplitudes of the defected metamaterial (Fig. \ref{f:harmonic}(f)). In the last configuration, transverse motion is spatially localized around the defected unit cell. Additionally, the blue rectangles in Fig. \ref{f:harmonic}(e, f) depict the voltage output of each unit cell. Since the displacement and voltage are evaluated at the excitation frequency laying inside the band gap range ($\Omega=6.01\times 10^{-3}$), negligible voltage output is observed for the periodic metamaterial. In Fig. \ref{f:harmonic}(f) the largest mechanical amplitude is observed at the defect (unit cell number 11), whereas the maximum voltage output is obtained from unit cells number 10 and 12. The voltage output of the $j$-th unit cell can be expressed as $v_j = {\bf{\Theta}_j^t\, \bf{u}\, \Omega^2}/({\Omega_{n_j}^2 - \Omega^2})$, where $\bf{\Theta}_j^t$ is the ($1\times m$) vector corresponding to the $j$-th line of the matrix $\bf{\Theta}^t$. Therefore, the low voltage output from 11\textsuperscript{th} unit cell is due to the mismatch of the excitation frequency and the electrical resonance of its attachment. In contrast, unit cells number 10 and 12 are excited around their electrical resonances, and thus, high voltage outputs are observed. Specifically, $v_{11}$ corresponds to 2.5\% of the total voltage output (summation of voltage output from each unit cell), whereas $v_{10}$ (or $v_{12}$) corresponds to 38.8\% of the total output.


\subsection{Space-time wave localization with programmable defects}\label{ss:time}

\begin{figure}[hbt]
	\centering
	\includegraphics[]{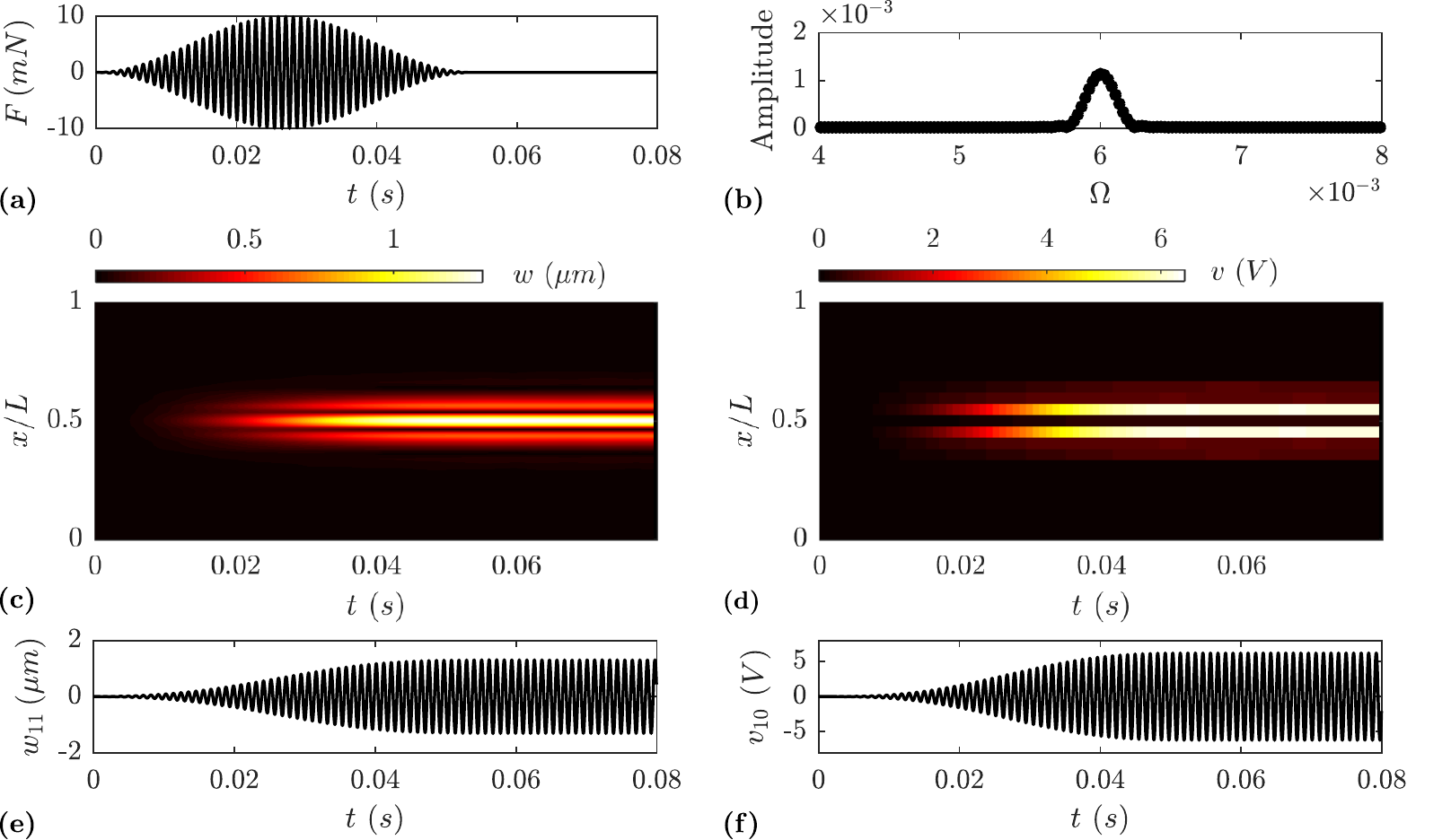}
	\caption{Sine burst excitation at $x/L = 0.5$ (a) and its frequency content (b). Envelope of the time-domain response along the beam with a single defect (c, d). Displacement at the middle of the cell with defect $w(0.5L)$ (e), and voltage output in the adjacent cell (f).}
	\label{f:time1}
\end{figure}

In this section, the time response of the piezoelectric metastructure with 21 unit cells is investigated. First, the defect fixed (in space and time) at 11\textsuperscript{th} unit cell is discussed. In this case, Eq. \ref{e:gov-ns} is numerically integrated to obtain the system response. Later, the space-time wave localization due to programmable defects is demonstrated, when Eq. \ref{e:gov-q} is solved. In all cases, the Newmark method \cite{Newmark1959} is employed to obtain the transient numerical solutions, this time integration method has been used with $\gamma = 1/2$ and $\beta=1/4$, which is the same as assuming that $\ddot{\bf{u}} = (\ddot{\bf{u}}_{k+1}+\ddot{\bf{u}}_k)/2$, $\ddot{\bf{v}} = (\ddot{\bf{v}}_{k+1}+\ddot{\bf{v}}_k)/2$ and $\ddot{\bf{q}} = (\ddot{\bf{q}}_{k+1}+\ddot{\bf{q}}_k)/2$ within the interval $(t_k,\, t_{k+1})$ \cite{Petyt1990}. Moreover, a constant time-step of $1.5783\times 10^{-5}$ s is considered, which properly describes the time response for frequencies up to of $\Omega = 0.008$. In all cases, a point excitation is applied at the 11\textsuperscript{th} unit cell with a sine burst shape (sinusoidal envelope) and central frequency $\Omega=6.01\times 10^{-3}$ (the defect-mode frequency), magnitude 0.01 N and duration of 50 cycles as shown in Fig. \ref{f:time1}(a,b).

The envelope of the transverse mechanical displacement and the voltage output of each unit cell as a function of space and time are shown in Fig. \ref{f:time1}(c, d), respectively. Figure \ref{f:time1}(c) shows that mechanical energy is mostly localized at unit cell number 11 (defect) as well as in its vicinity (unit cells 10 and 12). Figure \ref{f:time1}(d) shows that the maximum voltage output is observed at unit cells 10 and 12, in agreement with the discussions presented in Section \ref{ss:spectral}. Moreover, Fig. \ref{f:time1}(e, f) displays the time domain response of the transverse displacement at the defected unit cell and the voltage output from unit cell 10, respectively.

Having discussed the behavior of a metamaterial beam with a fixed point defect, the metamaterial with programmable defects is now explored. Space-time wave localization is performed by the spatial-temporal modulation of a defect in the periodic array of electromechanical local resonators (i.e., the defect is programmable). While the unit cell with defect is gradually changed to the periodic array configuration by gradually tuning its electrical resonant frequency (i.e., electromechanical local resonator) to $\Omega_t$, the shunt circuit of the subsequent unit cell is simultaneously updated from the periodic configuration $\Omega_t$ to open circuit (or to a significantly larger inductance such that $\Omega_{n_j} = 2\Omega_t$). Since vibration energy is localized in the initial defect, we expect to transfer this energy between adjacent unit cells, and hence, to control the wave localization. 

The rate of change of the local resonator frequency, $\mathrm{d}\Omega_n/\mathrm{d}t$, plays a major role in the energy transfer over space and time. A perfect energy transition between subsequent unit cells, which is accomplished by a smooth and synchronized changing of the local resonator frequencies, is the key point for space-time wave localization by programmable defects. With a fast transition (i.e., non-smooth) strategy, the vibration energy will leak through the structure, blocking the wave localization. A discussion regarding the transition effects into the space-time wave localization is presented in Appendix \ref{ap:trans}. In this work, a cosine function is used to modulate the local resonator (i.e., shunt circuit) properties during the space-time transitions because its time derivative is zero at both beginning and end of the transition interval and $\mathrm{d}\Omega_n/\mathrm{d}t$ is continuous for every $t$. In addition, the transition speed and smoothness can be controlled by its total duration ($t_f-t_0$, defined in Eq. \ref{e:trans}). Therefore, the variation of the locally resonant frequency of the $j$-th unit cell is governed by the following law: 
\begin{subequations}\label{e:trans}
	\begin{align}
	& \frac{\Omega_{n_j}}{\Omega_t}(x,t) = g(t)\  \delta(x-(j-1/2)\Delta), \quad j = 1, 2, ..., n \text{, with}\\
	& g(t)	= 
	\begin{cases}
	\eta & \text{if } t < t_0\\
	\frac{1+\eta}{2}+\frac{|\eta-1|}{2} \cos\left( \frac{t-t_0}{t_f-t_0} \pi \right) & \text{if } t_0\leq t\leq t_f\\
	1 & \text{if } t > t_f\\	
	\end{cases},\quad \text{for tuning}, \label{e:trans-b}\\
	& g(t)	=
	\begin{cases}
	1 & \text{if } t < t_0\\
	\frac{1+\eta}{2}-\frac{|\eta-1|}{2} \cos\left( \frac{t-t_0}{t_f-t_0} \pi \right) & \text{if } t_0\leq t\leq t_f\\
	\eta & \text{if } t > t_f \label{e:trans-c}
	\end{cases},\quad \text{for detuning,} 
	\end{align}
\end{subequations}
where $\eta$ is a constant ($\eta\in \mathbb{R} ^+_*$); $t_0$ and $t_f$ are the initial and final time steps of the transition, respectively; and $\delta$ is the Dirac delta. The resonance of a local resonator varies from $\Omega_n/\Omega_t = 1$ (the periodic configuration) to $\Omega_n/\Omega_t = \eta$ (the defect configuration) and the transition between these two configurations is governed by the cosine function.
However, for low values of $|\eta-1|$, the defect mode is not properly characterized because its local resonator properties are closer to the other unit cells. \ref{ap:detuning} discusses the effect of detuning ratio ($\Omega_n/\Omega_t$) on the defect mode, which provides a minimum value around $\Omega_n/\Omega_t=1.5$ to properly characterize the defect. Consequently, $\eta=2$ is employed in all simulations along this work.

\begin{figure}[hbt]
	\centering
	\includegraphics[scale = 0.80]{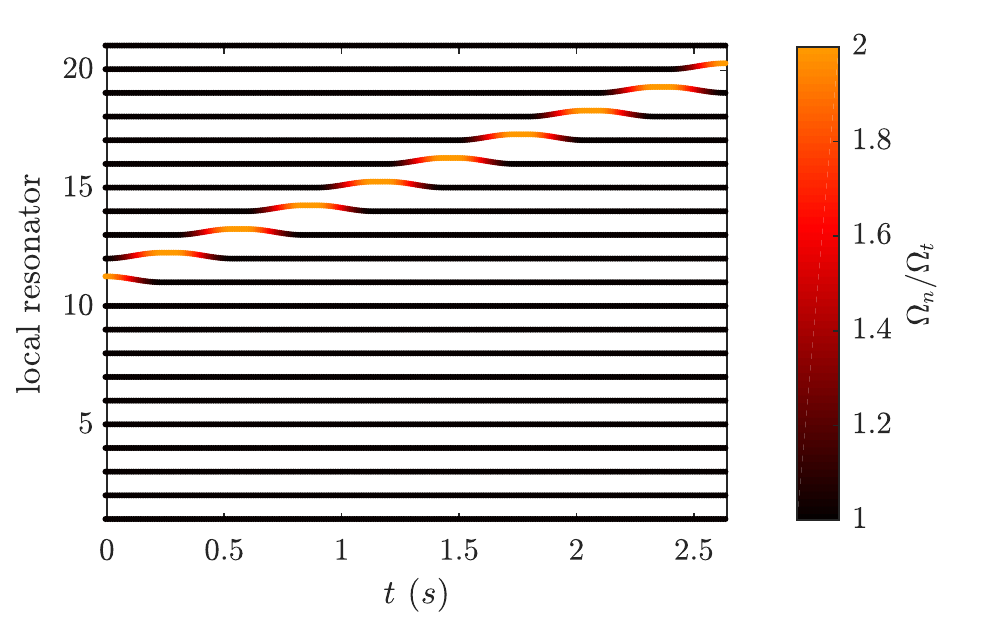}
	\caption{Example strategy for manipulating energy from unit cell 11 to 20 using \ref{e:trans} with $\eta=2$ depicted by a waterfall plot. The colors represent the detunig ratio: unit cell in the periodic configuration (black) and unit cell with defect (orange).}
	\label{f:waterfall}
\end{figure}

Figure \ref{f:waterfall} illustrates the use of Eq. \ref{e:trans} to control defects along the space and time in an electromechanical metamaterial with 21 unit cells. At $t=0$, the defect is placed at unit cell number 11 such that $\Omega_{n_{11}}/\Omega_t = 2$. The defect is gradually transferred from unit cell 11 to unit cell 12 when their electromechanical resonances are governed by Eq. \ref{e:trans} combined to Eq. \ref{e:trans-b} and Eq. \ref{e:trans-c}, respectively. In a more general case, in order to control the defect along space and time from unit cell 11 to unit cell 20, Eq. \ref{e:trans-b} is applied to a unit cell $j$ while Eq. \ref{e:trans-c} governs unit cell $j+1$, for $j = 11, 12, 13, ..., 19$.
 
\begin{figure}[h!]
	\centering
	\includegraphics[scale=0.95]{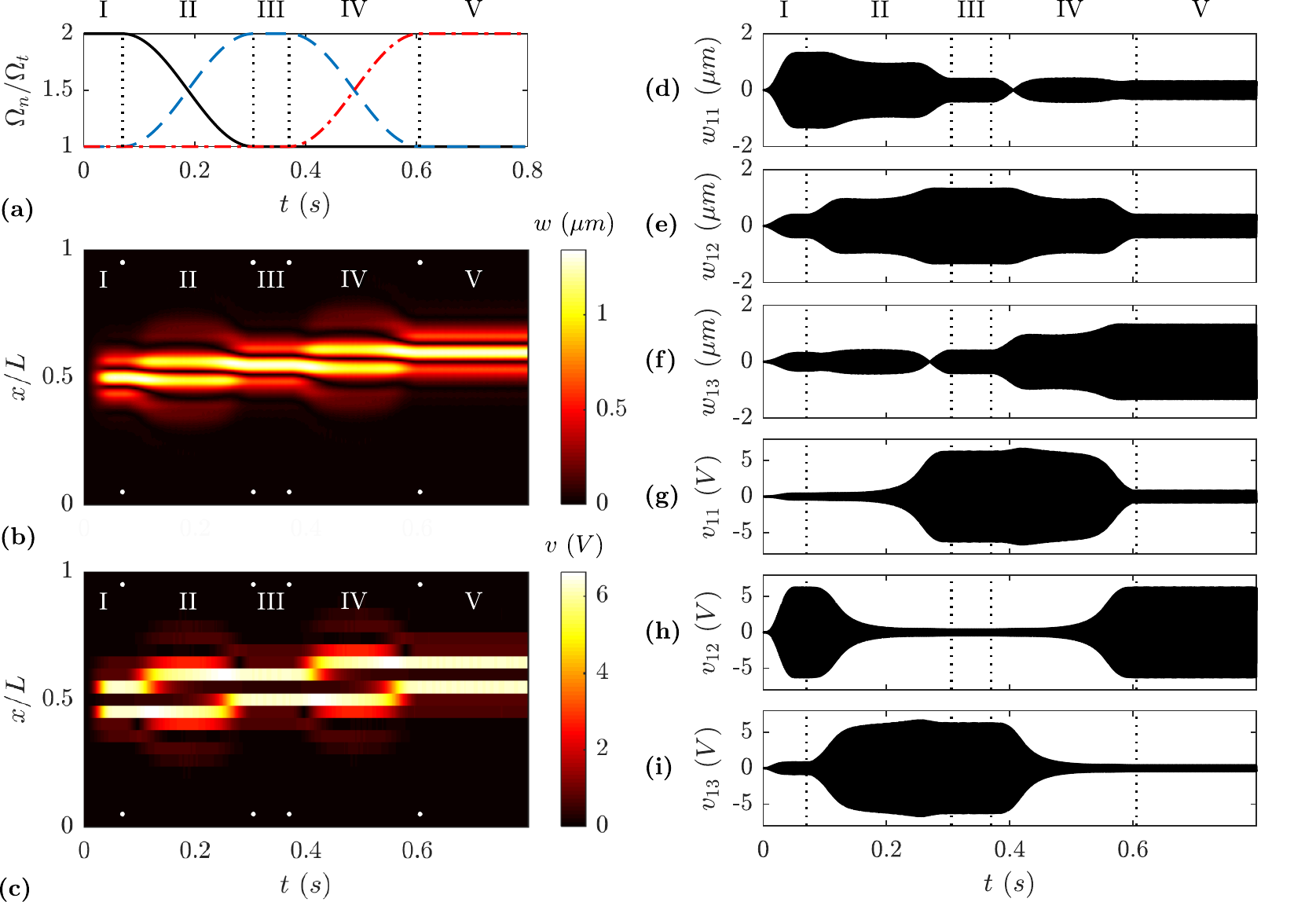}
	\caption{Variation of the local resonator frequencies of unit cells 11 (continuous black), 12 (dashed blue) and 13 (dash-dotted red) (a). Envelope of the time response along the beam with a moving defect (b, c). Displacement at the middle of unit cells 11, 12 and 13 ($x/L$ = 1/2, 23/42 and 25/42), respectively (d-f). Voltage output for cells 11, 12 and 13, respectively (g-i).}
	\label{f:time2}
\end{figure}

By using the transition strategy previously discussed, space-time wave localization along the electromechanical metamaterial beam due to programmable defects is performed. Again, a point defect is initially assumed at the center of the beam (unit cell number 11), where the excitation signal of Fig. \ref{f:time1}(a) is applied. Equation \ref{e:trans} is used to create a programmable defect (from unit cell 11 to 12, and then to 13), resulting in the transitions shown in Fig. \ref{f:time2}(a) and, then, the space-time wave transport and localization. The envelope of the transverse displacement along the beam and of the voltage output from each unit cell are shown in Fig. \ref{f:time2}(b, c), where the white dots delimits the stages I to V defined in Fig. \ref{f:time2}(a). To further illustrate, Fig. \ref{f:time2}(d-i) shows the displacement at the middle of unit cells 11, 12 and 13, as well as their voltage outputs. An animation of the system response is available in the supplementary material. 

During stage I, unit cell 11 is a point defect since it is tuned to $2\Omega_t$ ($\eta=2$) and all the other unit cells are tuned to $\Omega_t$. Therefore, vibration energy is confined inside and in the vicinity of the unit cell 11, as can be observed in Fig. \ref{f:time2}(b, d-f). Maximum voltage output is observed in unit cells 10 and 12 (Fig. \ref{f:time2}(c, g-i)), in agreement with the discussion of Section \ref{ss:spectral}. At the beginning of stage II (from 0.07 s to 0.16 s), vibration amplitude at unit cell 11 is almost constant (Fig. \ref{f:time2}(d)) since $\Omega_{n_{11}}/\Omega_t$ is slightly smaller than 2, while voltage output is low and constant (Fig. \ref{f:time2}(g)). Simultaneously, the rate of change of mechanical amplitude at unit cell 12 is high (Fig. \ref{f:time2}(e)) mostly due to energy flow from its own electrical attachment that is gradually detuned. Voltage output decreases in unit cell 12 (Fig. \ref{f:time2}(h)) due to unit cell detuning while voltage output at unit cell 13 increases. From 0.16 s to 0.25 s, vibration amplitude at unit cells 11 and 12 is quite similar, as observed in Fig. \ref{f:time2}(d, e) while voltage drops in unit cell 12 and increases in unit cell 13 (Fig. \ref{f:time2}(h, i)). In the end of stage II, the amplitude of mechanical oscillations drops in unit cell 11 since it becomes tuned to $\Omega_t$. Simultaneously, the amplitude of mechanical oscillations reaches a maximum in unit cell 12 since it becomes a defect. Furthermore, voltage output increases in unit cells 11 and 13, while reaches a minimum in unit cell 12.
 
As depicted in Fig. \ref{f:time2}(a), stage III characterizes a defect in unit cell 12 while all other unit cells are tuned to $\Omega_t$. Figure \ref{f:time2}(b, d-f) show that vibration energy has been transferred to unit cell 12, while low mechanical amplitude is observed in unit cells 11 and 13, whose voltage output is high. Thus, space-time wave localization (from unit cell 11 to 12) due to programmable defects was performed during stages I to III. The behavior for stages III to V is quite similar to stages I to III and is omitted for brevity. In the end of the process of Fig. \ref{f:time2}(a), the vibration energy is properly transferred and localized from unit cell 11 to unit cell 13, where the final defect position is. Therefore, we have demonstrated that the proposed strategy can be used to transfer and localize vibration energy between any two points along the metastructure domain.

\section{Space-time wave localization on a two-dimensional lattice metamaterial}\label{s:lattice}

\begin{figure}[hbt]
	\centering
	\includegraphics[scale = 0.85]{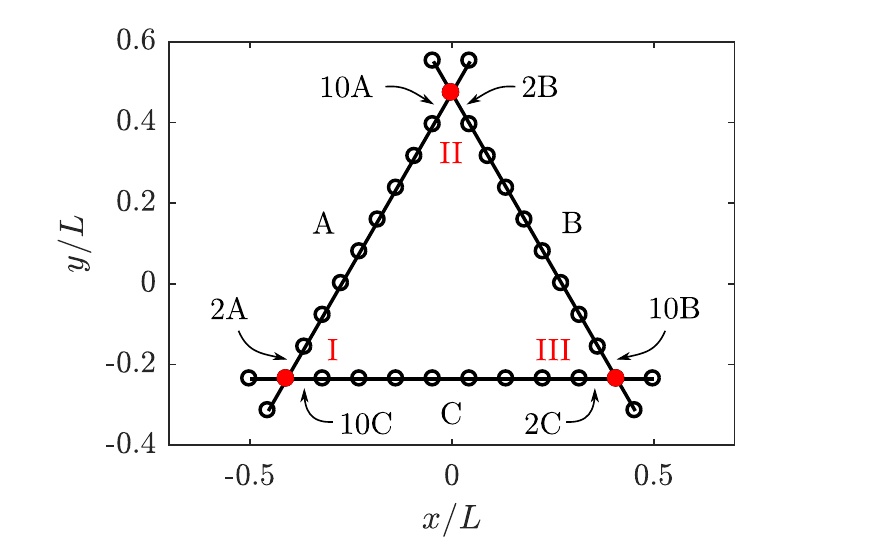}
	\caption{Triangular lattice structure constructed by the combination of three identical electromechanical bimorph beams with 11 unit cells and total length $L$. The open black circles indicate the unit cell boundaries while the filled red circles represent the nodes whose transverse motions are coupled. The beams are labeled with capital letters (A, B and C), the vertices with Roman numbers (I, II and III) and the unit cells with Arabic numbers (from 1 to 11) such that 2X and 10X indicate the second and next-to-last unit cells of beam X, respectively.}
	\label{f:triangle}
\end{figure}

The space-time wave localization concepts discussed in the previous section can be extended to more complex systems. Designing lattice metamaterials with programmable properties paves the way to smart, adaptive and versatile metamaterials, which can provide solutions to different problems including, for instance, reconfigurable wave localization in two-dimensional domains. The proposed lattice is a regular triangle whose edges are the piezoelectric metamaterial beams discussed in the previous sections. Although each piezoelectric beam has 11 unit cells, one unit cell is left outside the triangle in each vertex (as depicted in Fig. \ref{f:triangle}) in order to guarantee periodicity in the transition zones \cite{Pang2014,Jo2020}. Therefore, at the vertex, the unit cell with defect is surrounded at its boundaries for at least one unit cell in the periodic configuration (i.e., local resonator frequency tuned to $\Omega_t$), which creates the necessary band gap wall protection for the defect mode. In Fig. \ref{f:triangle}, the vertices are labeled as I, II and III, and the beams as A, B and C. The unit cells of each beam are referenced by their number and the respective beam literal, e.g., the second cell of beam A is cell 2A. At each vertex, the transverse motion of the beams are coupled by a high stiffness translational spring, such that the coupled nodes have the same transverse displacement; on the other hand, the rotations are not coupled. Although the cases explored in this section examine one single triangle, the results can be extended to more general configurations that combine several triangles in a hexagonal pattern. 

\begin{figure}[hbt]
	\centering
	\includegraphics[]{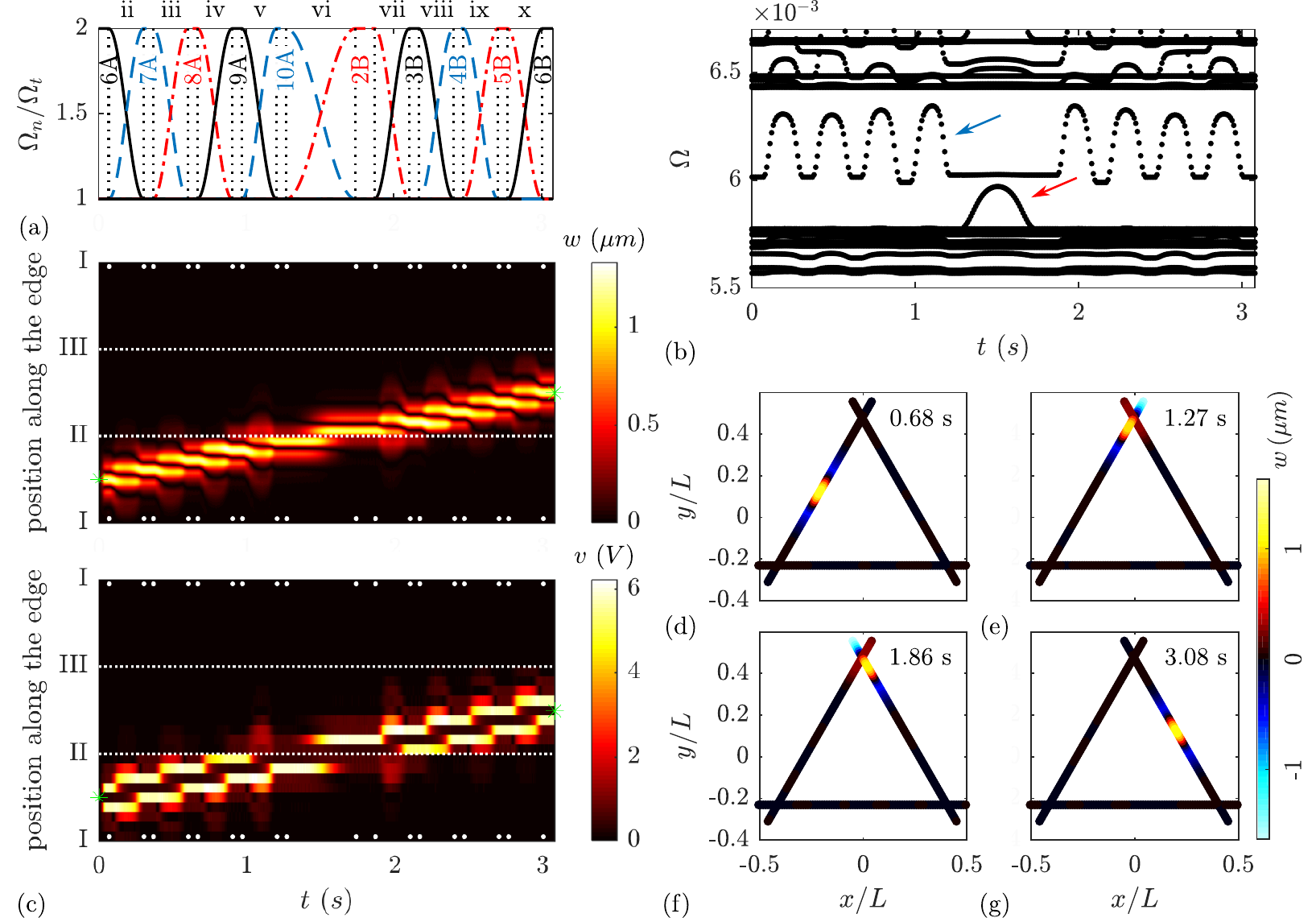}
	\caption{Variation of the electromechanical local resonant frequencies of selected unit cells for the first lattice case. Numbers and letters specify the unit cell that is being modified while Roman numbers indicate the stage label (a). Eigenvalues of the triangular lattice structure changing as the dynamic properties of the system changes. The blue arrow indicates the original defect mode and the red arrow shows the propagating mode emerging from the bulk (b). Envelope of the time-domain response for the triangular lattice structure when the defect is displaced from beam A to beam B (c). The vibration shapes of the structure for selected instants of time (d).}
	\label{f:time3}
\end{figure}

Figure \ref{f:time3} displays the time response results of the space-time wave localization by programmable defects in the periodic triangle. The excitation is due to a point force (Fig. \ref{f:time1}(a)) at the center of the beam A (unit cell 6A), where the defect is initially placed. Figure \ref{f:time3}(a) depicts the electromechanical local resonance transitions, governed by Eq. \ref{e:trans}, resulting in programmable defects from the center of beam A (unit cell 6A) to the center of beam B (unit cell 6B) passing through vertex II and, therefore, inducing space-time wave localization. Figure \ref{f:time3}(c) presents the envelope of the transverse displacements and voltage output along the circuit (from unit cell 6A to 6B) as a function of space and time, obtained from Eq. \ref{e:gov-q}. The response of the unit cells outside the lattice (i.e., cells 1A, 11A, 1B, 11B, 1C and 11C) are not included in Fig. \ref{f:time3} for clarity. The phenomenon is further illustrated in Fig. \ref{f:time3}(d), which shows the displacement field of the structure when the defect is localized  at cells 6A, 10A, 2B and 6B. An animated version of Fig. \ref{f:time3}(d) is also available in the supplementary material. 

In stage i, the point defect is at unit cell 6A while all others are tuned to $\Omega_t$ (i.e., the periodic configuration). The wave is localized around the defect (Fig. \ref{f:time3}(a, d), 0.07 s) and high voltage output is observed in the neighboring cells (Fig. \ref{f:time3}(c)). In stages ii to v, by using the same procedure described in \ref{ss:time} (including the transition times), the defect is gradually moved from cell 6A to 10A by updating frequencies of consecutive electromechanical local resonators (according to Eq. \ref{e:trans}), resulting in space-time energy transfer (Fig. \ref{f:time3}(c)). The same discussion of Section \ref{ss:time} holds for stages ii to v. In the end of stage v, the vibration energy is localized in the unit cell 10A. Relatively high amplitude is also observed in unit cell 11A since it is placed at the free end of the beam (Fig. \ref{f:time3}(d), 1.27 s). Stage vi includes the transition from beam A to beam B, or from unit cell 10A to 2B. Long lasting transition is required between the cells located at the vertex since the impedance is locally modified due to the beam coupling through the translational spring. Finally, in stages vii to xi, the energy is gradually transferred from unit cell 2B to 6B, following the same principles used in stages ii to v. In the end, the vibration energy is only localized at and in the vicinity of the final defect (unit cell 6B). 

Figure \ref{f:time3}(b) shows the system eigenvalues as a function of time as a consequence of the transitions depicted in Fig. \ref{f:time3}(a). Although eigenvalues are generally time-invariant, this analysis is still representative of the phenomenon since the rate of change of the inductance matrix is very small compared to the the angular frequencies analyzed, leading to Eq. \ref{e:gov-ns} with quasi-constant coefficients. This figure evidences two distinct mechanisms. The first one is observed when the defect is transitioning between two consecutive unit cells in the same beam ($0<t<1.20$ s and $1.88<t<3.08$ s). The defect-mode frequency gradually shifts between $\Omega=6.01\times10^{-3}$ (when a single cell is detuned) and $\Omega=6.30\times10^{-3}$ (when $\Omega_n=1.5\Omega_t$ for two adjacent cells). On the other hand, when the defect is transitioning from one beam to the other ($1.20<t<1.88$ s), the defect-mode frequency remains almost constant and a propagating mode (red arrow in the figure) emerges from the lower band gap limit. For small transition times, the defect wave mode gets contributions from the bulk (i.e., propagating) wave modes and, therefore, the vibration energy is leaked along the structure (please, check Appendix \ref{ap:trans} for more details about the fast transition issues). For an appropriated vertex transition in Fig. \ref{f:time3}(a), 440 cycles was settled with a total time of 471 ms.

\begin{figure}[h!]
	\centering
	\includegraphics[]{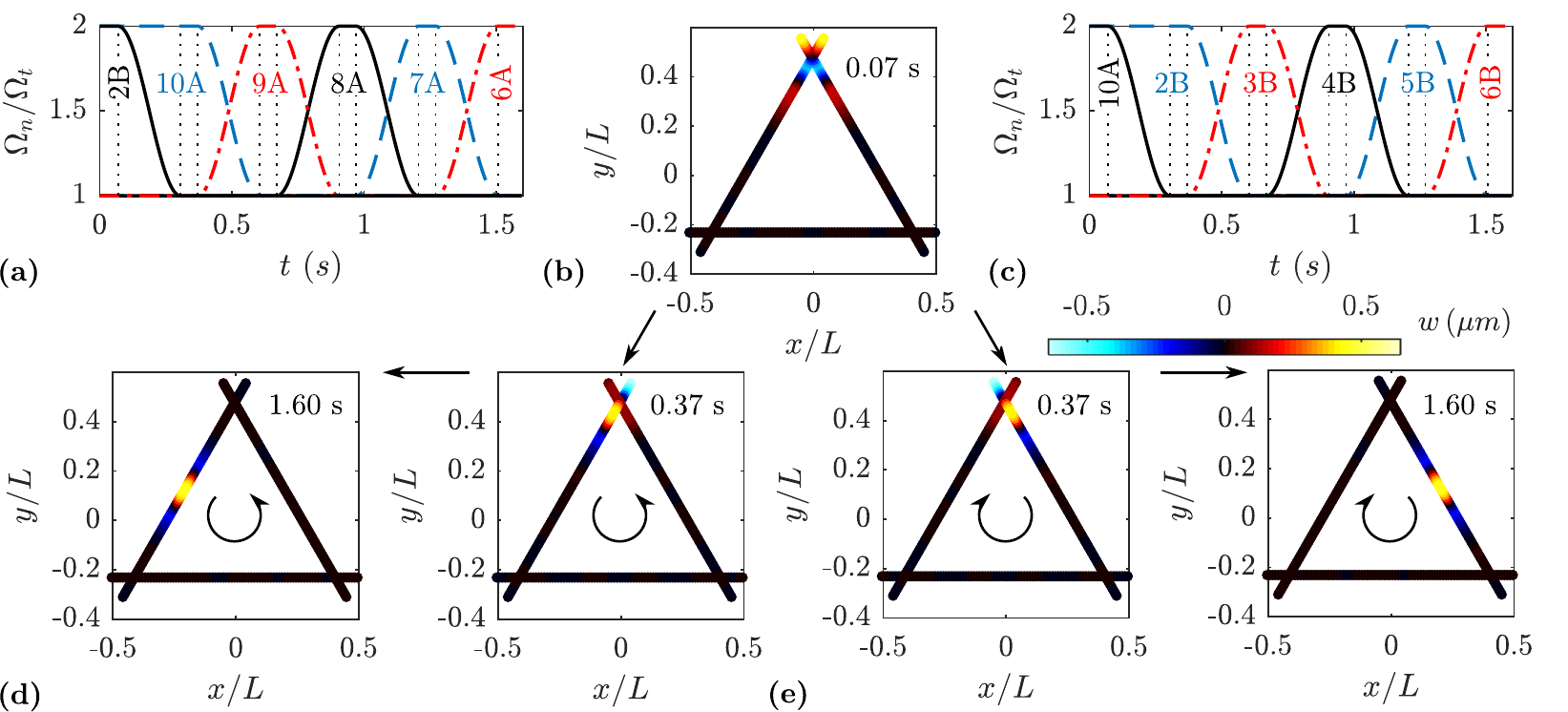}
	\caption{Variation of the electromechanical locally resonant frequencies of selected unit cells for the second lattice case (a, c). Displacement field of the structure for selected instants of time: energy confined in vertex II (b), transferring and confining energy into beam A (d), and into beam B (e).}
	\label{f:time4}
\end{figure}

Another case using the triangular configuration is presented in Fig. \ref{f:time4}. The point load excitation depicted in Fig. \ref{f:time1}(a, b) is now applied at vertex II. As can be observed in Fig. \ref{f:time4}(a, c), point defects are placed at unit cells 2B and 10A, consequently, the vibration is localized around vertex II. First, the defect is gradually moved along beam A according to the transitions presented in Fig. \ref{f:time4}(a) and, therefore, the wave originally localized at the vertex II is relocated to unit cell 6A (Fig. \ref{f:time4}(b, d)). Moreover, considering the transitions in Fig. \ref{f:time4}(c), the vibration energy is transferred from the vertex II to unit cell 6B (Fig. \ref{f:time4}(b, e)). Therefore, programmable defects can be used in the lattice to confine energy in a certain edge of the triangle and also in a specific point of this edge. Furthermore, if vertex II is assumed as an input port, the lattice metamaterial can be programmed to use vertex I and/or vertex III as output ports (or any different combination) and, this way, space-time wave localization could be performed along any desired circuit and direction of this lattice metamaterial with programmable defects.

\section{Conclusions}\label{s:conclusion}
 
Phononic crystals and metamaterials with point and line defects (i.e., waveguide circuits), whose wave mode lies inside the band gap, have been used to localize and to propagate waves.
In contrast to early configurations with fixed properties, recent investigations report tunable defects and waveguides that allows reconfigurable wave localization and reconfigurable wave guiding, among other applications. In this work, a novel programmable piezoelectric  metamaterial, which can confine vibration energy through point defects as well as transfer vibration energy along the structure by strategically programming the point defect in space and time, was presented. The point defect was introduced in the periodic array of electromechanical local resonators by tuning the electrical resonance of a unit cell (i.e. a local resonator) to a value significantly different from the other ones.

First, a piezoelectric metamaterial with a fixed point defect was investigated. Its band structure with coupled mechanical and electrical domains (i.e., multiphysics) was computed using the supercell technique, which reveled a flat defect band within the locally resonant band gap as well as a wave localization in the defective unit cell. This vibration localization was also observed in the harmonic response of the correspondent finite structure. Later, a strategy to control the defects along space and time was presented, which enabled a perfect energy transfer between unit cells, and hence, the wave localized in one unit cell could be perfectly transferred to any unit cell of the metastructure. This concept was also applied in a regular triangle whose edges are the electromechanical metamaterial beams (i.e., a two-dimensional unit cell lattice), which can be easily combined to other triangles in a hexagonal array for space-time wave localization along different two-dimensional circuits. 

This work opens new possibilities on wave localization (i.e., energy confinement) and manipulation (i.e., energy transferring) by using a smart electromechanical metastructures with programmable defects. Moreover, these concepts can be applied to wave devices with guiding, sensing, monitoring and harvesting capabilities. Finally, natural extensions of this work may involve investigations in two-dimensional continuum structures as well as the experimental realization of the proposed programmable metamaterial.


\section*{Acknowledgments}
The authors gratefully acknowledge the support of the Brazilian National Council for Scientific and Technological Development (CNPq) through Grants 308690/2019-2, 433456/2018-3 and 130603/2019-8. The authors also acknowledge the support of the Sao Paulo Research Foundation (FAPESP) through project reference numbers 2018/18774-6 (postdoctorate fellowship) and 2018/15894-0 (research project ``Periodic structure design and optimization for enhanced vibroacoustic performance: ENVIBRO'').



\bibliographystyle{unsrt}
\bibliography{reference}


\appendix


\section{Effects of the detuning ratio on the defect-mode frequency}\label{ap:detuning}

\begin{figure}[h!]
	\centering
	\includegraphics[]{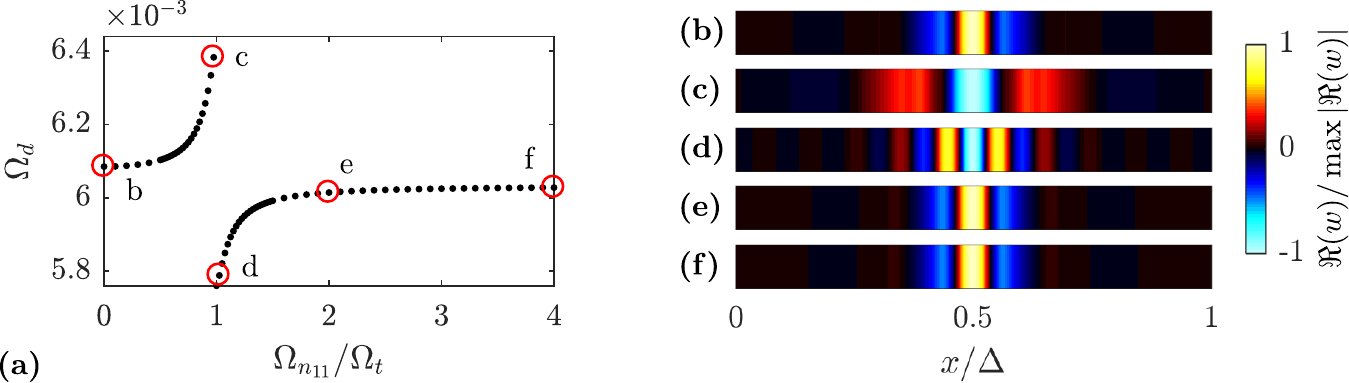}
	\caption{Detuning effect of the shunt circuit in the 11\textsuperscript{th} unit cell on the defect-band frequency (a). The limits of the ordinate axis (i.e., $\Omega_d = [5.76 \quad 6.44] \times 10^{-3}$) correspond to the band gap limits shown in Fig. 2. Wave mode shapes at $\kappa\Delta/\pi=1$ and at $\Omega_{n_{11}} /\Omega_t =$ 0 (b), 0.975 (c), 1.025 (d), 2 (e) and 4 (f), which are indicated by red circles in (a).}
	\label{f:Od_r}
\end{figure}

The effect of the local resonator detuning ($\Omega_n/\Omega_t$) on the defect-mode frequency is shown in \ref{f:Od_r}. Considering the electromechanical metamaterial beam with 21 unit cells, the defect is placed on the 11\textsuperscript{th} unit cell by changing the inductance of its shunt circuit. The supercell technique described in \ref{ss:spectral} is used to compute the band structure for each different configuration, and hence, to perform the parametric analysis of defect mode. 

When the detuning ratio $\Omega_{n_{11}}/\Omega_t$ is one, the local resonator of 11\textsuperscript{th} unit cell is tuned to $\Omega_t$ as all others unit cells of the supercell (i.e., the periodic configuration) and, therefore, the frequency branches reach the limits of the band gap (with frequencies presented in Fig. \ref{f:bs}(a)). For detuning ratios smaller than one, the defect emerges from the upper band gap limit until it reaches the defect-band frequency of $\Omega=6.08\times10^{-3}$ when the ratio is closer to zero (point $b$ in Fig. \ref{f:Od_r}(a)). Under this condition, the shunt circuit of the unit cell 11 would work, for instance, as a short-circuit. On the other hand, for detuning ratios larger than 1, the branch emerges from the lower band gap limit towards the defect-band frequency of $\Omega=6.03\times10^{-3}$ as the ratio increases. The defect-band frequencies are slightly different  when $\Omega_{n_{11}}/\Omega_t\rightarrow 0$ and $\Omega_{n_{11}}/\Omega_t\rightarrow \infty$  due to the effective rigidity of the piezoelectric material \cite{Hagood1991,Thorp2001}. For instance, the effective Young's modulus of a piezoelectric material connected to a resonant shunt circuit with  resonant frequency $\Omega_n$ is
\begin{equation}\label{key}
Y_p^{eff} = \frac{Y_p}{1-k_{31}^2} \left( 1 - \frac{k_{31}^2}{1-\Omega^2/\Omega_n^2} \right),
\end{equation}
where $k_{31}=d_{31}/\sqrt{s_{11}^E \epsilon_{33}^T}$ is the electromechanical coupling coefficient. Therefore, the piezoelectric material is more rigid for $\Omega_n/\Omega_t\rightarrow \infty$ than for $\Omega_n/\Omega_t\rightarrow0$ and, consequently, the corresponding defect-band frequencies are distinct. Fig. \ref{f:Od_r} also shows the wave modes for points $b$ to $f$ in Fig. \ref{f:Od_r}(a). When $\Omega_{n_{11}}/\Omega_t$ is closer to 1, the wave modes of Fig. \ref{f:Od_r}(c, d) show vibration energy is spread along the beam while for greater values of $|\Omega_{n_{11}}/\Omega_t-1|$ (Fig. \ref{f:Od_r}(b, e, f)) the energy is localized inside and in the vicinity of the defect. Therefore, $\Omega_n/\Omega_t$ has to be sufficiently different from 1 to obtain a defect. Accordingly, $\Omega_n/\Omega_t=2$ is a reasonable choice to characterize a defected unit cell in the metamaterial (Fig. \ref{f:Od_r}(e)).

\section{Effects of transition time on the space-time wave localization}\label{ap:trans}

In this appendix, the effects of different time transitions (governed by Eq. \ref{e:trans}) on the space-time wave localization is discussed. For this purpose, the metastructure with 21 unit cells and a defect in the 11\textsuperscript{th} unit cell is considered, this system was presented in \ref{ss:time}. The excitation is due to the point load depicted in \ref{f:time1} applied at the defect and Eq. \ref{e:gov-q} is solved, in this section, using a time-step of $1.1837\times 10^{-5}$ s, which properly describes the time-domain response for frequencies up to $\Omega = 0.012$. The effect of transition times (namely, 220, 65, 10 and 0 (switch) cycles of $\Omega=6.01\times 10^{-3}$) on the localized wave mode transferring from unit cell 11 to 12 is investigated.

\begin{figure}[h!]
	\centering
	\includegraphics[scale=1]{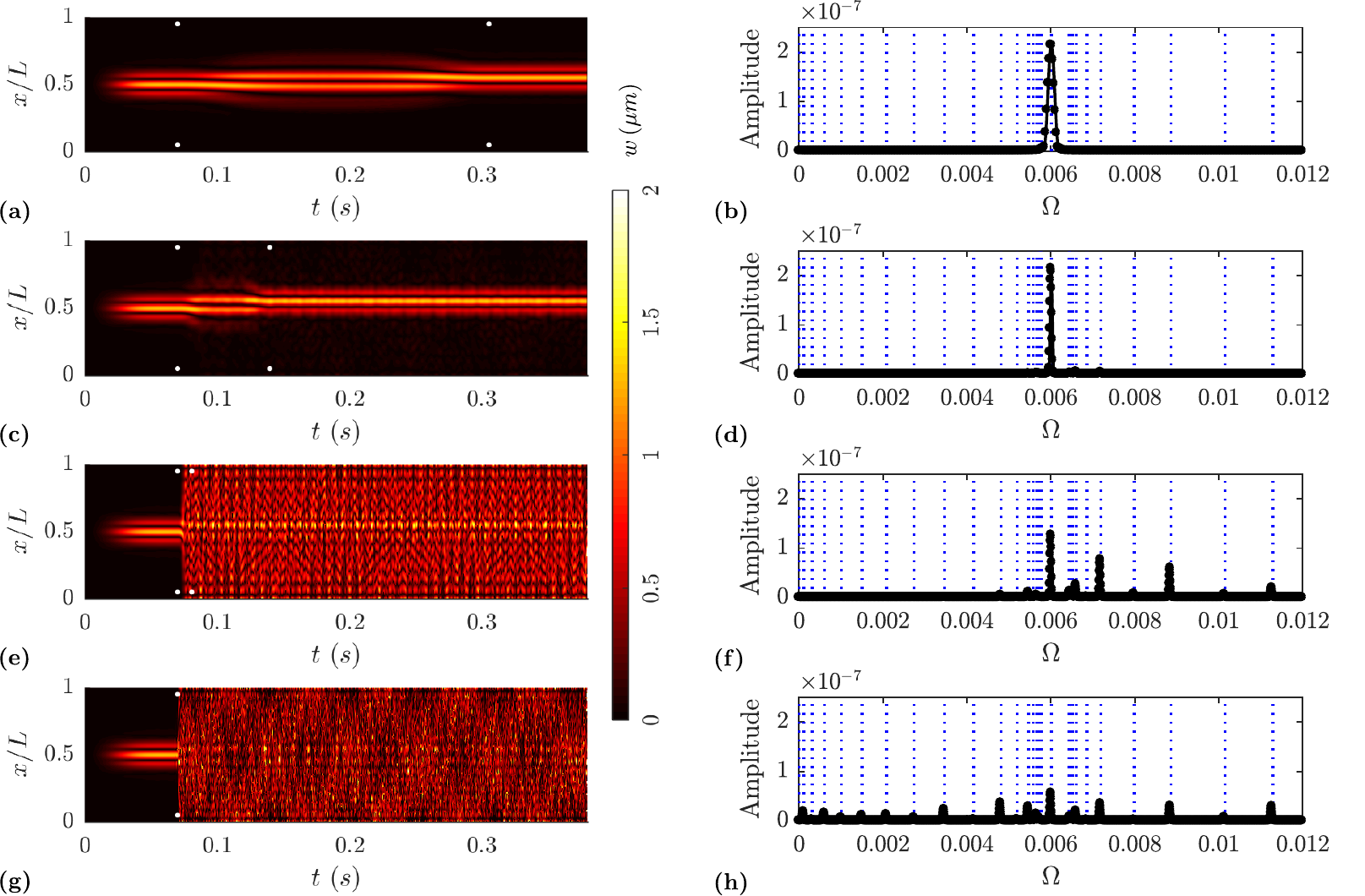}
	\caption{System response for transition time of 220 (a, b), 65 (c, d), 10 (e, f) and 0 (g, h) cycles of $\Omega=6.01\times 10^{-3}$. Envelope of the displacement field along the structure (a, c, e, g), where the white dots indicate the start and end of the transition. Discrete Fourier transform of the displacement at the middle of unit cell 12 ($x/L=23/42)$ evaluated after the end of the transitions (b, d, f, h), where the blue dotted lines indicate the eigenvalues of the system.}
	\label{f:ap}
\end{figure}

Figure \ref{f:ap}(a, c, e, g) shows the envelope of the displacement field as function of space and time, and Fig. \ref{f:ap}(b, d, f, h) displays the discrete Fourier transform (DFT) of the displacement at the middle of unit cell 12. For the DFT computation, the interval of time between the end of the transitions to the end of the simulation is considered; in addition, a Hanning window is applied to the signal before the DFT computation. The blue dotted lines in Fig. \ref{f:ap}(b, d, f, h) show the eigenvalues of the system at the final time (i.e., configuration with defect on unit cell 12). For slow transitions, space-time wave localization (from unit cell 11 to 12) is properly performed (Fig. \ref{f:ap}(a)) and the beam vibrates exclusively at the defect-mode frequency (Fig. \ref{f:ap}(b)). A similar behavior is observed for the 65-cycle transition, although vibration energy is slightly spread beyond the vicinity of unit cell 12 (Fig. \ref{f:ap}(c)) and negligible contributions from bulk modes are observed (Fig. \ref{f:ap}(d)). On the other hand, when the frequency of the local resonator abruptly changes, such as in the 10- and 0-cycle transitions, the variation of the dynamic properties (i.e., inductance) is too fast to allow the vibration energy transfer between consecutive unit cells whose inductances are being modified, and hence, the wave localization transference does not occur. As consequence, the vibration energy, instead of passing from one unit cell to the other, spreads throughout the beam (Fig. \ref{f:ap}(e, g)) exciting several natural vibration modes. Moreover, the spectrum band of excitation modes becomes wider as the rate of change of the local resonator properties increases (Fig. \ref{f:ap}(f, h)). Therefore, the variation of the local resonator properties must be sufficiently slow to achieve proper space-time wave localization, as in Fig. \ref{f:ap}(a, b).

\end{document}